\documentclass[journal]{journal}

\ifCLASSINFOpdf
\else
\fi
\usepackage{graphicx}
\usepackage{multirow} 
\usepackage[ruled,linesnumbered]{algorithm2e}
\usepackage{amsmath}
\usepackage{booktabs}
\usepackage{hyperref}
\usepackage{caption}

\begin{document}
\title{DCI: A Coordinated Allocation and Filling Workload-Aware Dual-Cache Allocation GNN Inference Acceleration System}
%\title{DCI: A Workload-Aware Dual-Cache System for GNN Inference with Adjacency Matrix and Node Feature Cache}

\author{Yi~Luo, Yaobin Wang${^\ast}$, Qi~Wang, Yingchen~Song, Huan~Wu, Qingfeng~Wang, and Jun~Huang      
	\thanks{Yi Luo, Qi Wang, Yingchen Song, and Huan Wu are students with Southwest University of Science and Technology, Mianyang, China. Emails: yiluose@gmail.com, 86972190@qq.com, yingchensong@foxmail.com, huanwu0713@foxmail.com.}
	\thanks{Yaobin Wang is the corresponding author and is a faculty member with Southwest University of Science and Technology, Mianyang, China. Email: wangyaobin@foxmail.com.}
	\thanks{Qingfeng Wang and Jun Huang are faculty members with Southwest University of Science and Technology, Mianyang, China. Emails: 475914518@qq.com, huangjuncs@swust.edu.cn.}}

\maketitle
\thispagestyle{empty}

\begin{abstract}

Graph Neural Networks (GNNs) are powerful tools for processing graph-structured data, increasingly used for large-scale real-world graphs via sampling-based inference methods. However, inherent characteristics of neighbor sampling lead to redundant data loading during GNN inference, compounded by inefficient data transfers between host and GPU memory, resulting in slow inference and low resource utilization. Existing methods to accelerate GNN inference face several challenges: (1) low practical GPU memory utilization, (2) overlooking adjacency matrix locality, and (3) long preprocessing time. To address these challenges, we introduce DCI, an efficient workload-aware dual-cache allocation system for GNN inference acceleration. DCI allocates cache capacities for both node features and adjacency matrices based on workload patterns during the pre-sampling phase, leveraging a lightweight cache-filling algorithm to optimize data loading efficiency. Experimental results demonstrate that DCI accelerates sampling and node feature loading, achieving end-to-end inference speedups of 1.18$\times$ to 11.26$\times$ compared to DGL, and 1.14$\times$ to 13.68$\times$ over RAIN, while reducing preprocessing time by 52.8\% to 98.7\%. Additionally, DCI outperforms state-of-the-art single-cache inference systems by achieving speedup of 1.08$\times$ to 1.32$\times$. We also compared DCI with DUCATI's dual-cache population strategy. Our lightweight population algorithm allows DCI to achieve nearly the same inference speed while keeping preprocessing time to less than 20\% of that required by DUCATI.
	
\end{abstract}


\begin{IEEEkeywords}
Graph Neural Networks, dual-cache, large graph, inference.
\end{IEEEkeywords}

% \fi
\IEEEpeerreviewmaketitle

\section{Introduction}
\IEEEPARstart{G}{raphs}, as non-Euclidean data, effectively capture complex relationships between entities and are widely used in real-world applications. Graph Neural Networks (GNNs) have achieved significant success in tasks like vertex classification and link prediction ~\cite{reau2023deeprank,wu2020comprehensive,jiang2022graph,Liu2022A}. A graph, composed of nodes and edges representing entities and their relationships, provides a structural representation of these connections. However, with the advent of the Big Data era, real-world graphs are often enormous and grow rapidly. For instance, the Ogbn-papers100M dataset~\cite{hu2020open} contains 111 million vertices and 1.6 billion edges, with an adjacency matrix and node features totaling around 70GB, given their size, full-graph inference for GNNs is often impractical due to CPU and GPU memory constraints. To address this, sampling-based mini-batch training methods~\cite{liu2022gnnsampler,chiang2019cluster,qiu2020gcc,zeng2019graphsaint} have been developed, which generate subgraphs through stochastic sampling. This approach effectively reduces memory usage while maintaining high predictive accuracy, making it a practical solution for handling large-scale graphs.

GNN inference plays a critical role in deploying trained models in real-world applications, yet performing inference on large-scale graphs remains time-consuming. While substantial research has focused on accelerating GNN inference, most efforts have centered around channel pruning~\cite{zhang2022accelerating,yik2022input} and model distillation~\cite{wang2023graph,gao2022efficient}, both of which require model re-training. Cache-based approaches~\cite{cai2023dsp,liu2024efficient,zhang2021pcgraph,lin2020pagraph} aim to mitigate CPU-GPU data transfers by caching frequently accessed node features in GPU memory. Additionally, the use of unified virtual addressing (UVA) has been proposed~\cite{min2021large} to enhance the processing of irregular data accesses during GNN training.

\begin{figure}[b]
	\centering
	{\includegraphics[width=0.58\textwidth, trim=2.4cm 11.6cm 1cm 10.1cm, clip]{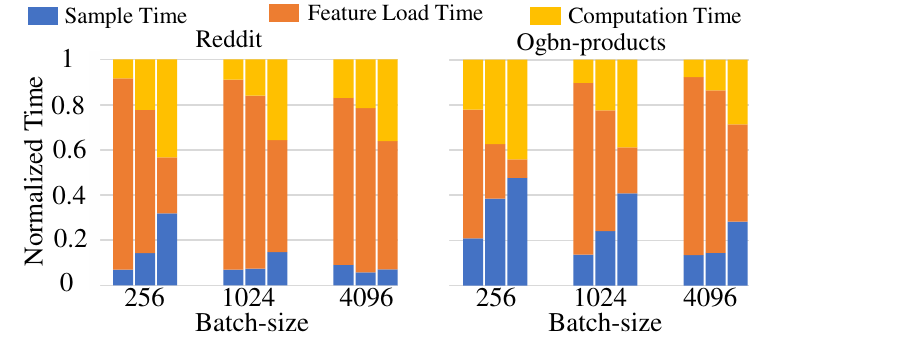}}
	\caption{Decomposition of total time for performing inference across different datasets, with specified left-to-right fan-out: `2,2,2', `8,4,2’, and `15,10,5'.}
	\label{fig1}
\end{figure}

As shown in Fig.~\ref{fig1}, through inference experiments using the GraphSAGE model on two real-world graphs (Reddit~\cite{hamilton2017inductive} and Ogbn-products~\cite{hu2020open}, where a complete inference on the test set is performed through sampling-based methods), It was observed that mini-batch preparation time (the sum of sampling and node feature loading time) accounts for 56\%-92\% of the total inference time. Furthermore, current cache-based systems ~\cite{liu2024efficient,zhang2021pcgraph,lin2020pagraph} are built on the fundamental assumption that feature loading is more time-consuming than sampling. However, this assumption may not always hold in practice. As illustrated in Fig.~\ref{fig1}, the proportion of sampling and feature loading times varies, indicating that simple node feature caching is not the optimal solution.

\begin{figure}[t]
	\centering
	{\includegraphics[width=0.49\textwidth, trim=2.7cm 15.5cm 4.6cm 6.2cm, clip]{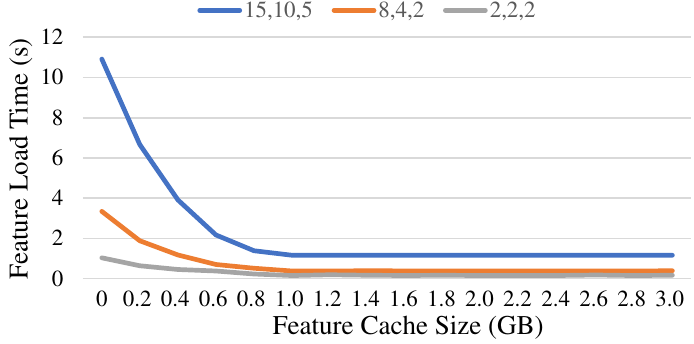}}
	\caption{Impact of node feature caching on reducing node feature loading time. Experimental results were obtained using GraphSAGE on the Ogbn-Products dataset with different fan-out, with a batch size of 4096.}
	\label{figa1}
\end{figure}

Different cache capacities were allocated for node features, and the inference results under varying capacities are shown in Fig.~\ref{figa1}. It was found that GraphSAGE does not benefit from a cache capacity greater than 1GB. Therefore, using all idle GPU memory for node feature caching leads to low effective GPU memory utilization due to the long tail effect, where a small number of high-frequency samples dominate while low-frequency samples, which are also cached, contribute little to performance. To address the low GPU resource utilization in existing GNN inference, an adjacency matrix cache is introduced together with the node feature cache, forming a dual-cache inference system. This system allocates cache capacity for node features and adjacency matrices based on workload-awareness and a lightweight cache allocation algorithm, thereby accelerating both sampling and node feature loading processes and improving GNN inference efficiency.

This work makes the following contributions:
\begin{itemize}
	\item[$\bullet$] The GNN inference process is decomposed, and it is found that the preparation time of mini-batches occupies 56\%-92\% of the total GNN inference time. Additionally, the time proportions of the two stages, sampling and node feature selection, vary significantly, highlighting the limitations of existing cache-based GNN inference systems. 
	
	\item[$\bullet$] A dual-cache system for GNN inference is proposed, combining node feature and adjacency matrix caching, along with an efficient workload-aware cache allocation strategy that optimizes GPU memory usage and balances the caching requirements of node features and adjacency matrices. DCI introduces a lightweight cache-filling algorithm that effectively reduces preprocessing overhead, improves GPU utilization, and accelerates inference speed.

	\item[$\bullet$] All experiments were conducted on an NVIDIA GeForce RTX 4090 GPU. The approach outperforms DGL, RAIN, and state-of-the-art single-cache inference systems, achieving up to 13.68$\times$ speedup. Compared to RAIN, preprocessing time was reduced by 52.8\% to 98.7\%. Compared to DUCATI’s dual-cache population strategy, DCI achieved at least a 81.38\% reduction in preprocessing time, while maintaining nearly identical inference performance.
\end{itemize}

\begin{table}[t]
	\centering
	\caption{Summary of sampling statistics for the Ogbn-products dataset.}\label{tabbb1}
	\begin{tabular}{c|c|c|c|c}
		\hline
		\multicolumn{2}{c|}{\textbf{Hyperparameter}} & \textbf{Test-} & \textbf{Loaded-} & \multirow{2}{*}{\centering \textbf{Load/Test}} \\ \cline{1-2} 
		\textit{Batch size} & \textit{fan outs} & \textbf{nodes} & \textbf{nodes} & \\ \hline
		& 15,10,5 &  & 1,030,270,033 &  465.534 \\ \cline{2-2} \cline{4-5}
		256 & 8,4,2 &  & 203,853,530 & 92.113 \\  \cline{2-2} \cline{4-5}
		& 2,2,2 &  & 47,989,922 & 21.685 \\ \cline{1-2} \cline{4-5} 
		& 15,10,5 &  & 851,864,912 & 384.921 \\ \cline{2-2} \cline{4-5} 
		1024 & 8,4,2 & 2,213,091 & 193,778,584 & 87.560 \\  \cline{2-2} \cline{4-5} 
		& 2,2,2 &  & 47,306,640 & 21.376 \\ \cline{1-2} \cline{4-5} 
		& 15,10,5 &  & 531,357,988 & 240.098 \\ \cline{2-2} \cline{4-5} 
		4096 & 8,4,2 &  & 165,620,769 & 74.837 \\  \cline{2-2} \cline{4-5} 
		& 2,2,2 &  & 44,914,351 & 20.295 \\ \hline 
	\end{tabular}
\end{table}

\section{Background and Related Works}
This section first introduces the background of GNNs and the compressed sparse column (CSC) format for sparse matrices, followed by a description of sampled GNN inference and the work that accelerates GNN inference.

\subsection{Graph Neural Networks}
This work targets attributed graphs, where vertices or edges are associated with a large number of features in addition to the structural information of the graph. A GNN model usually consists of multiple layers~\cite{lin2020pagraph}, within the same layer, all vertices share the same aggregation neural network and transformation neural network, the computation between different layers follows the traditional iterative processing model of vertex-centred graphs, at each layer, each vertex transforms the features from its neighbours by aggregating them, and then transforms these features into output features using a neural network, and these output features will be used as the input features are passed on to the next layer~\cite{hamilton2017inductive}, the output of the last layer can be used for tasks such as node classification and link prediction~\cite{huang2018adaptive,yu2023pn,liu2024rt}.

\begin{figure}[b]
	\centering
	{\includegraphics[width=0.5\textwidth, trim=0.5cm 0.6cm 0.3cm 0.4cm, clip]{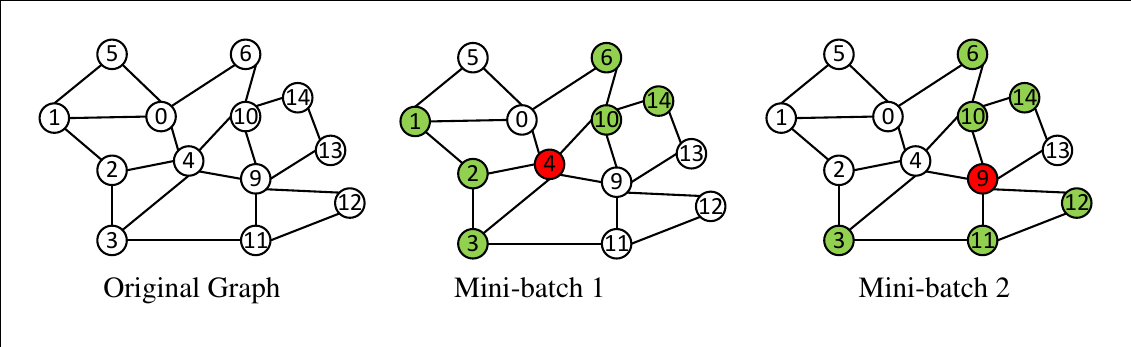}}
	\caption{Selection of mini-batches during the entire inference process.}
	\label{figa2}
\end{figure}

\subsection{Sampling-based Inference with GPU}
Since CPUs are slow in handling massively parallel tasks, sampling-based GNN inference usually requires transferring data to GPU to take advantage of their powerful parallel computing capability to accelerate the inference process. Due to GPU memory constraints, loading the entire graph onto the GPU is impractical for large graphs. To solve this problem, sampling has been widely adopted as a typical optimization solution~\cite{chen2018fastgcn, chen2017stochastic}, where neighbourhood sampling-based inference selects mini-batch based on the given batch size and fan-out, and inputs the mini-batch into the model for inference. The computational cost is greatly reduced while achieving almost the same accuracy.

\begin{figure}[t]
	\centering
	{\includegraphics[width=0.45\textwidth, trim=5cm 2.2cm 3cm 1.65cm, clip]{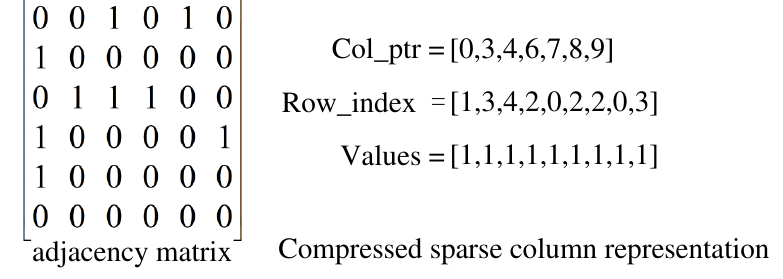}}
	\caption{Adjacency matrix in CSC format.}
	\label{fig4}
\end{figure}

However, GNN inference based on neighbor sampling requires selecting multiple mini-batches, and the same nodes may be selected across different mini-batches. As shown in Fig.~\ref{figa2}, both mini-batch 1 and mini-batch 2 select nodes 3, 6, 10, and 14, leading to redundant data loading when these mini-batches are loaded onto the GPU, resulting in significant time overhead. This phenomenon is further confirmed by experiments on the Reddit and Ogbn-products datasets, as shown in Table~\ref{tabbb1}. Each batch size corresponds to three fan-out values. The smaller the batch size, the greater the number of batches, consequently increasing the likelihood of sampling the same nodes across different batches. In the worst-case scenario, this results in up to 465.534$\times$ redundant data loading.

\subsection{The Storage of the Graph Dataset}
Graph datasets typically contain two main pieces of information, the adjacency matrix and the node features, where the node features are stored as compact 2D tensor and the adjacency matrix is usually stored by the COO, CSR and CSC formats~\cite{bulucc2009parallel}. The CSC format is the most suitable format for sampling because the sampling process requires fast access to the in-neighbours of the target node, so modern GNN systems~\cite{fey2019fast,wang2019deep} usually use a compressed sparse column format to store the adjacency matrix. As shown in Fig.~\ref{fig4}, CSC uses three arrays to store the adjacency matrix information, the Col\_ptr array contains the starting offset position of the first element of each column, the Row\_index array contains the row indices corresponding to the elements in the Values array, and the Values array contains all the non-zero elements in the matrix.

\subsection{Related Works}
Existing work related to GNN inference focuses on channel pruning~\cite{yik2022input,zhang2022accelerating} and model distillation~\cite{gao2022efficient,wang2023graph}, as well as some cache-based work~\cite{zhang2021pcgraph,liu2024efficient}.

\textbf{Work based on channel pruning.} J. Yik et al.~\cite{yik2022input} proposed a method for pruning the input features, which reduces the amount of raw data processed by the model to reduce the communication overhead between CPU and GPU, while greedy and regression-based algorithms are developed to determine which features to retain for optimal prediction accuracy. W. Zhang et al.~\cite{zhang2022accelerating} proposed a soft-channel pruning method with a ladder pruning pattern. This method reduces the computation on unimportant graph node features and achieves performance acceleration, while preserving the inference accuracy of GNNs.

\textbf{Work based on model distillation.} X. Gao et al.~\cite{gao2022efficient} proposed a new adaptive propagation order method that generates a personalised propagation order based on the topological information of each node, which is capable of avoiding redundant computation and allows for a flexible trade-off between accuracy and speed. W. Zhang et al.~\cite{wang2023graph} proposed a graph explicit neural network
(GENN) framework, which aims to solve the problem of MPNNs’ over-reliance
on over-reliance on node features and high inference latency, this approach alleviates the dependence on node features and improves the efficiency and accuracy
of inference.

\textbf{Cache-based work.} Cache-based inference systems for GNNs are relatively rare. L. Zhang et al.~\cite{zhang2021pcgraph} proposed PCGraph, which supports adaptive GNN inference and feature partition caching. By partitioning target vertices and sequentially caching their corresponding partitions, PCGraph reduces redundant data transfer between CPU and GPU and significantly decreases vertex embedding computation time through adaptive inference techniques. T. Liu et al.~\cite{liu2024efficient} introduced RAIN, an efficient GNN inference system based on locality-sensitive hashing (LSH), which clusters similar mini-batches and reuses node features across neighboring batches to minimize redundant data loading. 

In addition, there are several cache-based GNN training systems. Z. Lin et al.~\cite{lin2020pagraph} proposed PaGraph, the first system to utilize idle GPU memory for storing node features. PaGraph's approach is based on the assumption that real-world graphs follow a power-law distribution, leading it to prioritize storing high-degree nodes. However, this assumption does not hold for all scenarios. To address this, A. Xin et al.~\cite{ai2023neutronorch} proposed NeutronOrch, which uses a hotness-aware, layer-based task orchestration method. NeutronOrch offloads the training tasks of frequently accessed vertices to the CPU while the GPU reuses their embeddings with bounded staleness. Additionally, Z. Xin et al.~\cite{zhang2023ducati} developed DUCATI, which adds an adjacency matrix cache (Adj-Cache) together with the traditional node feature cache (Nfeat-Cache) to further accelerate the GNN training process.

The above channel pruning and model distillation based efforts require retraining the model, and the cache based efforts mainly use the free memory of the GPU to cache frequently accessed node information, essentially exploiting the locality of node features. UVA technology was introduced in DGL(V0.8.1)~\cite{wang2019deep}, which allows GPUs and CPUs to share the same virtual address space, allowing for more efficient data transfers, where GPUs and CPUs have direct access to each other's memory space. However, the current UVA-based approach does not take advantage of the locality of the data.

\begin{figure*}[!t]
	\centering
	\includegraphics[width=0.9\textwidth, trim=0.5cm 0.6cm 0.3cm 0.4cm, clip]{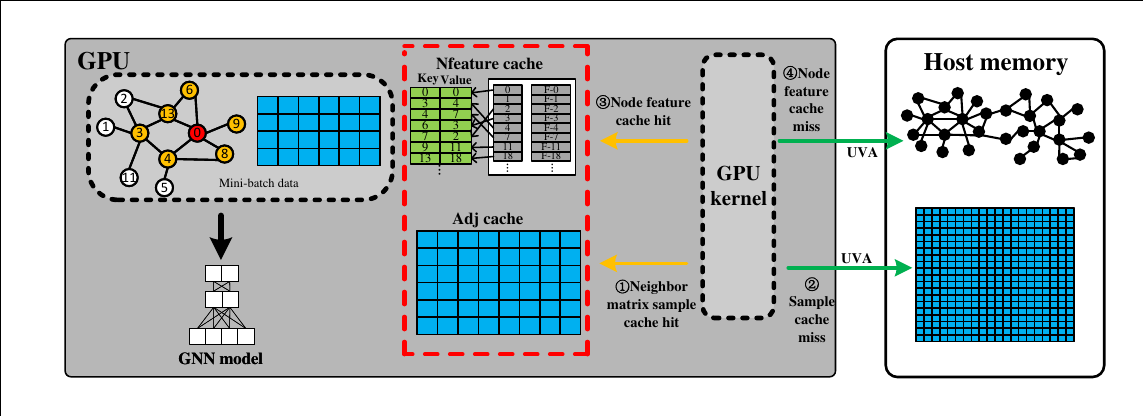}
	\caption{Overall framework of DCI.} 
	\label{fig5}
\end{figure*}

\section{Motivation}
Experiments on two real-world graphs (Ogbn-products and Reddit) show that mini-batch preparation time accounts for 56\%-92\% of the total GNN inference time, and the preparation time is inversely proportional to the batch size. The inference process was decomposed, revealing that the loads of the sampling and feature collection stages are imbalanced. Consequently, existing cache-based acceleration methods for inference have limitations, as they utilize all available GPU memory to store node features. Given that most real-world graphs follow a power-law distribution, caching only a small portion of the data can often yield good results. As shown in Fig.~\ref{figa1}, when conducting inference with GraphSAGE on the Ogbn-products dataset, it was observed that increasing the cache capacity beyond 1GB did not provide additional benefits. Analysis of the results identified redundant data access during the sampling and feature loading stages as the primary factor slowing down the entire inference process. Moreover, the current node feature caching systems fail to fully utilize GPU resources, as using all available memory to cache node features results in inefficient memory usage. To address this, an adjacency matrix cache and a lightweight cache-filling algorithm were introduced to accelerate GNN inference.

\section{The Proposed Method: DCI}\label{DCI}

Inspired by the findings of prior experiments, the DCI system has been developed—a \underline{d}ual-\underline{c}ache system tailor-made for \underline{i}nference applications, featuring a lightweight allocation and filling strategy for cache capacity. This is the first instance where an adjacency matrix cache has been integrated into a GNN inference system, in conjunction with a node feature caching strategy. Additionally, an efficient dual-cache filling algorithm has been formulated that substantially improves the efficiency of preprocessing operations in the inference process for large-scale graphs, offering a solution that is considerably more lightweight compared to DUCATI.

The overall framework of DCI is shown in Fig.~\ref{fig5}. The idea of DCI is to sense the total capacity of GPU available for caching through workload and allocate the total capacity to node features and adjacency matrix for storing the adjacency matrix elements and node features that need to be accessed frequently during sampling process. If the cache hits during sampling and feature selection, the data is loaded directly from the GPU memory, and if the cache does not hit, the required data is loaded from the host memory by UVA technique, thus reducing the redundant loading of data during sampling and node feature loading in the inference process.

DCI's core optimisation is a cache capacity allocation algorithm that uses the available GPU memory for storing adjacency matrix elements and node features that are frequently accessed during sampling and node feature selection. A key issue arises: since there is no iterative operation in the inference phase of GNNs, the preprocessing time cannot be spread across multiple epochs as in training, i.e., DCI's cache capacity allocation and cache filling algorithms require lightweight approaches.

\subsection{Workload-Aware Cache Capacity Allocation Algorithm}
The algorithm is workload-aware because the memory consumption of the GPU does not vary significantly from batch to batch during sampling-based inference. The convention of previous work, as described in~\cite{lin2020pagraph,yang2022gnnlab}, is followed by running several batches of pre-sampling to predict the maximum load on the GPU's memory resources. Based on this, the available memory capacity of the GPU is determined, and the sampling and node feature loading times are computed, with caches for the adjacency matrix and node features allocated based on the ratio of these two times. It is worth noting that, since only a few pre-samplings were performed and completely accurate information about the workload could not be obtained, a portion of the GPU space must be reserved to avoid memory overflow errors. It has been shown through experiments that reserving 1GB of memory is completely sufficient. This is the same operation as in PaGraph~\cite{lin2020pagraph}, and although not all datasets require 1GB of space, it is used as a reference value for the experimental setup.

The allocation cache capacity is determined by Equation (1).
\begin{equation}
	\begin{aligned}
		C_{\text{adj}} &= \frac{\sum_{k=1}^{n} t_{\text{sample},k}}{\sum_{k=1}^{n} (t_{\text{sample},k} + t_{\text{feature},k})} \times C \\
		C_{\text{feat}} &= \frac{\sum_{k=1}^{n} t_{\text{feature},k}}{\sum_{k=1}^{n} (t_{\text{sample},k} + t_{\text{feature},k})} \times C	
	\end{aligned}
\end{equation}

\begin{figure*}[t]
	\centering
	\includegraphics[width=0.9\textwidth, trim=1.1cm 1cm 0.8cm 0.7cm, clip]{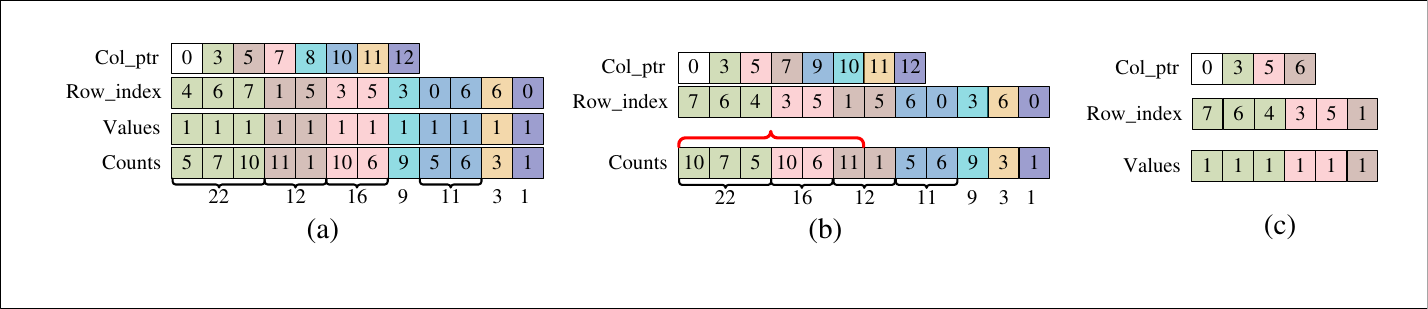}
	\caption{Caching process for the adjacency matrix.} 
	\label{fig6}
\end{figure*}

In Equation (1), \(C\) denotes the total cache capacity available to the GPU for caching neighborhood matrix elements and node features, \(T_{\text{sample}}\) represents the time occupied by sampling during the pre-sampling process, and \(T_{\text{feature}}\) represents the time occupied by feature loading. \(n\) denotes the number of preprocessing batches. \(C_{\text{adj}}\) and \(C_{\text{feat}}\) correspond to the cache capacities for the adjacency matrix and node features, respectively.

\subsection{Double Cache Filling Algorithm}
During the pre-sampling process, the number of visits to each node and each element in the neighbourhood matrix is also counted. A one-dimensional tensor is used to count the number of visits to a node, and the average number of visits to a node during the pre-sampling process is obtained. Instead of sorting the number of visits to a node, the nodes with a number of visits greater than the average are directly selected to populate their features into the node feature cache. If the feature cache still has capacity after filling all the node features with accesses greater than the average number of accesses, the node features with fewer accesses than the average are then filled. Inside the GPU, the node features are quickly located in the GPU memory through a hash table.

As shown in Fig.~\ref{fig6}(a), the modified CSC format includes the Counts array used to store the number of times each element has been accessed. Fig.~\ref{fig6}(b) shows the array sorted according to the number of accesses. Two levels of sorting have been implemented for the adjacency matrix. The first level sorts each node. For example, node 0 has three elements accessed 22 times, while node 1 has two elements accessed 12 times, so node 0 is placed before node 1. The second level sorts the elements within each node. For instance, node 0's elements (4, 6, and 7) are sorted by the number of accesses, resulting in the order 7, 6, 4. In Fig.~\ref{fig6}(b), the elements enclosed by braces are populated into the adjacency matrix cache, while those not enclosed are not populated due to insufficient cache capacity.

Fig.~\ref{fig6}(c) shows the CSC array filled into the adjacency matrix cache. At this point, the Counts array is deleted. For node 2, there were originally two elements, but now only one element is cached. In the sampling can be based on the original length and the size of the cache length to determine when the cache hit, for example, the sampling process want to go to access the nth element of node 2, the length of the cache is 1, if n is less than or equal to 1 then the cache hit, otherwise it is not hit, the details of the filling of the adjacency matrix is shown in Algorithm ~\ref{algorithm1}.

In Algorithm ~\ref{algorithm1}, line 1 calculates the storage volume of the CSC array, if its storage volume is less than or equal to the cache capacity then the CSC array is cached in its entirety, otherwise it goes to line 6 and starts to go to the total number of accesses to neighboring nodes by each node, lines 10 and 11 are sorted in descending order according to the total number of accesses and reorganize the CSC array, lines 12 to 15 sort the number of accesses to a node's neighbors and it is the ones with high accesses that the Neighbors are ranked first and finally fill the neighboring moment cache according to the cache capacity.

\begin{algorithm}[t]
	%	\small
	\caption{Adjacency Matrix Cache Filling Algorithm by DCI.}
	\label{algorithm1}
	\SetKwData{Left}{left}\SetKwData{This}{this}\SetKwData{Up}{up}
	\SetKwFunction{Union}{Union}\SetKwFunction{FindCompress}{FindCompress}
	\SetKwInOut{Input}{Input}\SetKwInOut{Output}{Output}
	
	\Input{$C_{\text{adj}}$, $\text{Col\_ptr}$, $\text{Row\_index}$, $\text{Values}$, and $\text{Count}$.}
	
	\Output{$\text{New\_col\_ptr}$, $\text{New\_row\_index}$, and $\text{New\_values}$.}
	\BlankLine
	%	//Update the short rows index phase\;
	$cache_{volume} \gets \text{computeCSCVolume}$
	
	\If{$cache_{volume} \leq C_{\text{adj}}$}{
		$\text{New\_Colptr}$, $\text{New\_Rowindex}$, $\text{New\_Values} \gets$ All of the CSC array
	}
	\Else{
		\emph{Initialize an array $node\_totals$ to store total visit counts for each node}\;
		\For{$i \gets 0$ \textbf{to} $\text{length}(sorted\_nodes) - 1$}{
			\emph{$node\_totals[i] \gets \sum(\text{Count}[\text{Col\_ptr}[i]:\text{Col\_ptr}[i+1]])$}\;
		}
		\emph{$sorted\_nodes \gets \text{argsort}(-node\_totals)$}\;
		\emph{Reorder $\text{Col\_ptr}$, $\text{Row\_index}$, and $\text{Values}$ according to $sorted\_nodes$}\;
		\For{$i \gets 0$ \textbf{to} $\text{length}(sorted\_nodes) - 1$}{
			\emph{$elements \gets \text{Count}[\text{Col\_ptr}[i]:\text{Col\_ptr}[i+1]]$}\;
			\emph{$sorted\_nodes \gets \text{argsort}(-node\_totals)$}\;
		}
		\emph{$\text{New\_Colptr}$, $\text{New\_Rowindex}$, $\text{New\_Values} \gets$ Slicing the CSC array}\;
	}
	\Return $\text{New\_col\_ptr}$, $\text{New\_row\_index}$, $\text{New\_values}$.
\end{algorithm}

\section{Experiment and Evaluation}
\subsection{Experiment Setup}
\textbf{Platform:} Experiments were conducted on a machine equipped with an Intel Core i9-13900KF CPU, 128GB of DDR4 RAM, and an NVIDIA GeForce RTX 4090 GPU (24GB memory). The system runs Ubuntu 20.04 and includes CUDA v11.8, DGL(v0.8)~\cite{wang2019deep}, and PyTorch(v2.1.2)~\cite{pytorch111}.

\begin{table*}
	\centering
	\caption{Dataset statistics (“m” stands for multi-class classification).}\label{tab1}
	\begin{tabular}{ccccccc}
		\toprule
		\textbf{Dataset} & \textbf{Nodes} & \textbf{Edges} & \textbf{Average degree} & \textbf{Feature} & \textbf{Classes} & \textbf{Train/Val/Test} \\
		\midrule
		Reddit & 232,965 & 11,606,919 & 50 & 602 & 41 & 0.66/0.10/0.24 \\ 
		Yelp & 716,480 & 6,977,410 & 10 & 300 & 100 (m) & 0.75/0.10/0.15 \\ 
		Amazon & 1,598,960 & 132,169,734 & 83 & 200 & 107 (m) & 0.85/0.05/0.10 \\ 
		Ogbn-products & 2,449,029 & 61,859,140 & 25 & 100 & 47 & 0.08/0.02/0.90 \\ 
		Ogbn-papers100M & 111,059,956 & 1,615,685,872 & 29.1 & 128 & 172 & 0.78/0.08/0.14 \\ 
		\bottomrule
	\end{tabular}
\end{table*}

\begin{figure*}[!t]	
	\centering	
	\includegraphics[width=\textwidth, trim=0.4cm 2.7cm 0.5cm 0cm, clip]{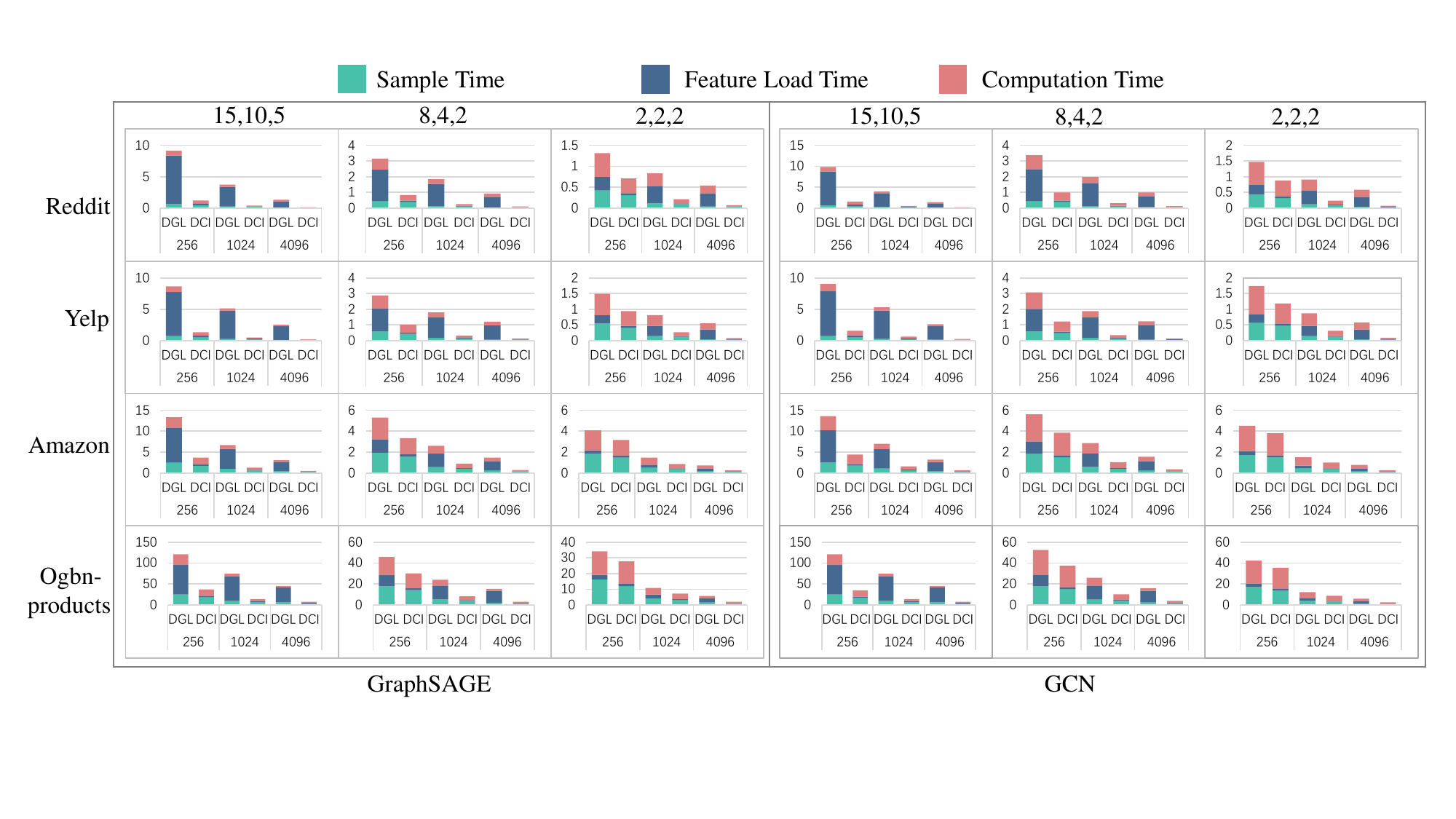}	
	\caption{DGL and DCI inference time for four datasets with different parameters (Y-axis unit: seconds, X-axis: batch size).} 	
	\label{figa3}
\end{figure*}

\begin{table}[htbp]
	\centering
	\setlength{\tabcolsep}{11pt}
	\caption{Model architectures.(FC: fully connected layer. Agg: type of aggregating operation. Hidden: hidden embedding dimension).}\label{taba1}
	\begin{tabular}{ccccl}
		\toprule
		Model &Layer  & Agg  & Allpy & Hidden \\ 
		\midrule
		GraphSAGE& 3 & sum & FC & 128\\
		GCN& 3 & avg & FC & 128 \\ 
		\bottomrule
	\end{tabular}
\end{table}

\textbf{Datasets:} For experimental evaluation, five widely used datasets were chosen, as shown in Table~\ref{taba1}. The Reddit~\cite{hamilton2017inductive} social network, a popular online forum, where posts are grouped into communities. The Yelp~\cite{zeng2019graphsaint} categorizes types of businesses based on customer reviews and friendships among users. The Amazon~\cite{zeng2019graphsaint} categorizes products based on buyers’ reviews and interactions. The Ogbn-products~\cite{hu2020open} represents the Amazon product co-purchase network, where nodes are products and edges indicate that they are frequently bought together, and the Ogbn-papers100M~\cite{hu2020open}  is a directed citation graph of 111 million papers indexed by MAG. In its node set, about 1.5 million are ARXIV papers.  The datasets used in this experiment follow the divisions of previous experiments.

\textbf{Baselines:} To demonstrate the effectiveness of DCI, DCI is compared with the following baselines:
\begin{enumerate}
	\item \textbf{DGL:} DGL reduces the GNN computational model to several general sparse tensor operations, adopts a frame-neutral design, and is an efficient and flexible graph neural network framework.
	\item \textbf{SCI:} The state-of-the-art single-cache inference (SCI) system is used, which disables the adjacency matrix cache in the DCI architecture. Other than this, SCI and DCI share the same architecture.
	\item \textbf{RAIN:} RAIN proposes an efficient GNN inference system by proposing a strategy that samples target nodes according to the size of their node degree, clusters similar batches by Local Sensitive Hashing (LSH), and sequentially performs inference on similar batches so that data can be reused between two batches.
	\item \textbf{DUCATI:} DUCATI is a dual-cache system that adaptively determines the optimal cache allocation. It formulates the cache-filling process as a variant of the knapsack problem, prioritizing nodes with the highest value (impact on speed-to-size ratio) to accelerate mini-batch preparation.
\end{enumerate}

\textbf{Models:} In the following experiments, representative graph neural network models, GraphSAGE~\cite{hamilton2017inductive} and Graph Convolutional Network (GCN)~\cite{kipf2016semi}, were used, with more details provided in Table~\ref{taba1}. The same training model parameters were used in DCI, DGL, and RAIN. In these experiments, neighbour sampling was used, while RAIN employed its unique adaptive sampling strategy. All results were obtained by averaging five runs.

\subsection{Overall Performance}
\textbf{Comparison with DGL.} DCI is initially compared with the original GNN inference method in DGL to demonstrate the effectiveness of the approach. As shown in Fig.~\ref{figa3}, DCI and DGL inference performance across various datasets and parameter combinations is illustrated. At this stage, preprocessing time is excluded because inference tasks are executed periodically, and the preprocessing process can be considered as an offline scenario.Overall, DCI achieves speedups ranging from 1.22$\times$ to 11.26$\times$ (average 4.92$\times$) with GraphSAGE and 1.18$\times$ to 9.07$\times$ (average 4.22$\times$) with GCN under different parameter configurations.

The inference process is broken down into three stages: sampling, feature loading, and computation. The focus is on optimizing the first two stages. In GraphSAGE, DCI reduces sampling time by 16.22\% to 54.43\% (average 29.42\%) and feature loading time by 59.76\% to 96.83\% (average 90.62\%). In GCN, it reduces sampling time by 13.62\% to 49.07\% (average 27.31\%) and feature loading time by 50.52\% to 96.78\% (average 90.90\%). It is observed that, under the same batch size, the performance improvement of the method is smaller when the fan-out is smaller. This is because, with smaller fan-out settings, the proportion of time spent on the sampling process becomes relatively larger compared to larger fan-out settings, which is also supported by the time breakdown analysis in Fig.~\ref{fig1}. According to Amdahl's Law, in such cases, the overall performance gain of DCI is limited, with speedups of only 1.18$\times$ under certain parameter configurations.

\begin{figure}[t]
	\centering	
	\includegraphics[width=0.5\textwidth, trim=13.8cm 6cm 6cm 4.5cm, clip]{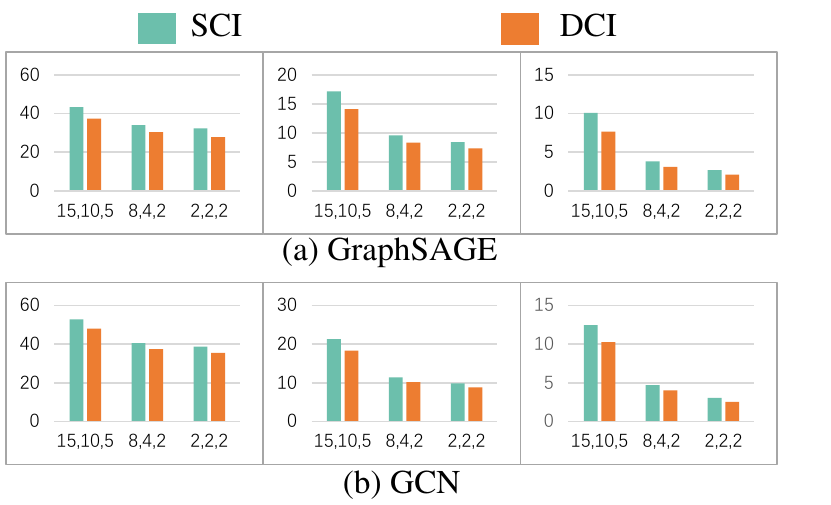}	
	\caption{Inference Time of SCI and DCI on the Ogbn-products Dataset Under Different Models and Parameter Settings (Y-axis unit: seconds, X-axis: batch size).} 	
	\label{figa4}
\end{figure}

\textbf{Comparison with SCI.} Previous experiments have validated the effectiveness of DCI over DGL's original inference method. To further assess the impact of adjacency matrix caching in DCI, it was compared with SCI across different models and parameters on the Ogbn-products dataset, as shown in Fig.~\ref{figa4}. DCI achieved speedups of 1.12$\times$ to 1.32$\times$ (average 1.20$\times$) in GraphSAGE and 1.08$\times$ to 1.22$\times$ (average 1.14$\times$) in GCN compared to SCI. Additionally, it demonstrates that DCI enhances GPU utilization, whereas single-cache systems underutilize memory, even when fully dedicating available space to feature storage.

\begin{table}[t]
	\centering
	\setlength{\tabcolsep}{5pt}
	\caption{Comparison of Preprocessing Time Between DCI and RAIN (BS: Batch Size; Products: Ogbn-products; Unit: s).}
	\label{tab4}
	\begin{tabular}{ccccccccc}
		\toprule
		\multirow{2}{*}{\centering \textbf{BS}} & \multicolumn{2}{c}{\textbf{Reddit}} & \multicolumn{2}{c}{\textbf{Yelp}} & \multicolumn{2}{c}{\textbf{Amazon}} & \multicolumn{2}{c}{\textbf{Products}}  \\ \cline{2-9}
		& \textit{RAIN} & \textit{DCI} & \textit{RAIN} & \textit{DCI} & \textit{RAIN} & \textit{DCI}& \textit{RAIN} & \textit{DCI}  \\ \hline
		256 & 5.05 & 0.26 & 5.23 & 0.40 & 15.17 & 0.55 & 31.43 & 0.42  \\ 
		1024 & 3.40  & 0.32 & 1.79 & 0.42 & 5.00 & 0.59 & 8.85 & 0.45 \\ 
		4096 & 3.41 & 0.32 & 0.96 & 0.45 & 3.76 & 0.72 & 4.92 & 0.66 \\
		\bottomrule
	\end{tabular}
\end{table}

\textbf{Comparison with RAIN.} DCI was compared with RAIN, which employs adaptive layer sampling. Following the parameter settings of the original authors of RAIN, the sampling layers were set to one, while DCI uses node-neighbor sampling with fan-out set to ‘15, 10, 5’, and the GraphSAGE model was employed. The comparison results are presented in Table~\ref{tab3}. During the experiments, it was observed that RAIN consumes a significant amount of GPU memory during inference. To test its scalability, the Ogbn-papers100M dataset was included, and the results show that RAIN encountered a RuntimeError: CUDA out of memory when trying to allocate 52.96 GB of GPU memory. Such substantial memory overhead severely limits the applicability of RAIN. In contrast, DCI successfully performed inference on the Ogbn-papers100M dataset using a single GPU (NVIDIA RTX 4090 24GB), demonstrating that DCI requires less hardware and is applicable in a wider range of scenarios.

\begin{table}[t]
	\centering
	\caption{Comparison of inference time between DCI and RAIN with different covariates on different datasets (Unit: seconds). }\label{tab3}
	\begin{tabular}{ccccc}
		\toprule
		{\textbf{Dataset}} & {\textbf{Batch size}} & {\textbf{RAIN}} & {\textbf{DCI}} & {\textbf{Speedup}} \\ \midrule
		& 256 & 5.59 & 1.23 & 4.56 \\ 
		Reddit& 1024 & 4.11 & 0.42 & 9.75  \\ 
		& 4096 & 2.12 & 0.16 & 13.03\\ \hline
		& 256 & 8.08 & 1.34 & 6.01 \\ 
		Yelp& 1024 & 4.21 & 0.51 & 8.19 \\
		& 4096 & 3.06 & 0.22 & 13.68 \\ \hline
		& 256 & 18.95 & 3.70 & 5.12 \\ 
		Amazon& 1024 & 8.30 & 1.31 & 6.34 \\ 
		& 4096 & 5.47 & 0.51 & 10.75 \\ \hline
		& 256 & 40.03 & 35.21 & 1.14 \\ 
		Ogbn-products& 1024 & 20.50 & 14.14 & 1.45 \\ 
		& 4096 & 18.81 & 7.65 & 2.46 \\ \hline
		& 256 & OOM & 19.76 & - \\ 
		Ogbn-papers100M& 1024 & OOM & 7.10 & - \\ 
		& 4096 & OOM & 3.71 & - \\
		\bottomrule
	\end{tabular}
\end{table}

\subsection{Preprocessing Overhead}

\begin{figure*}[!t]
	\centering	
	\includegraphics[width=\textwidth, trim=2.6cm 2.4cm 2.8cm 3.2cm, clip]{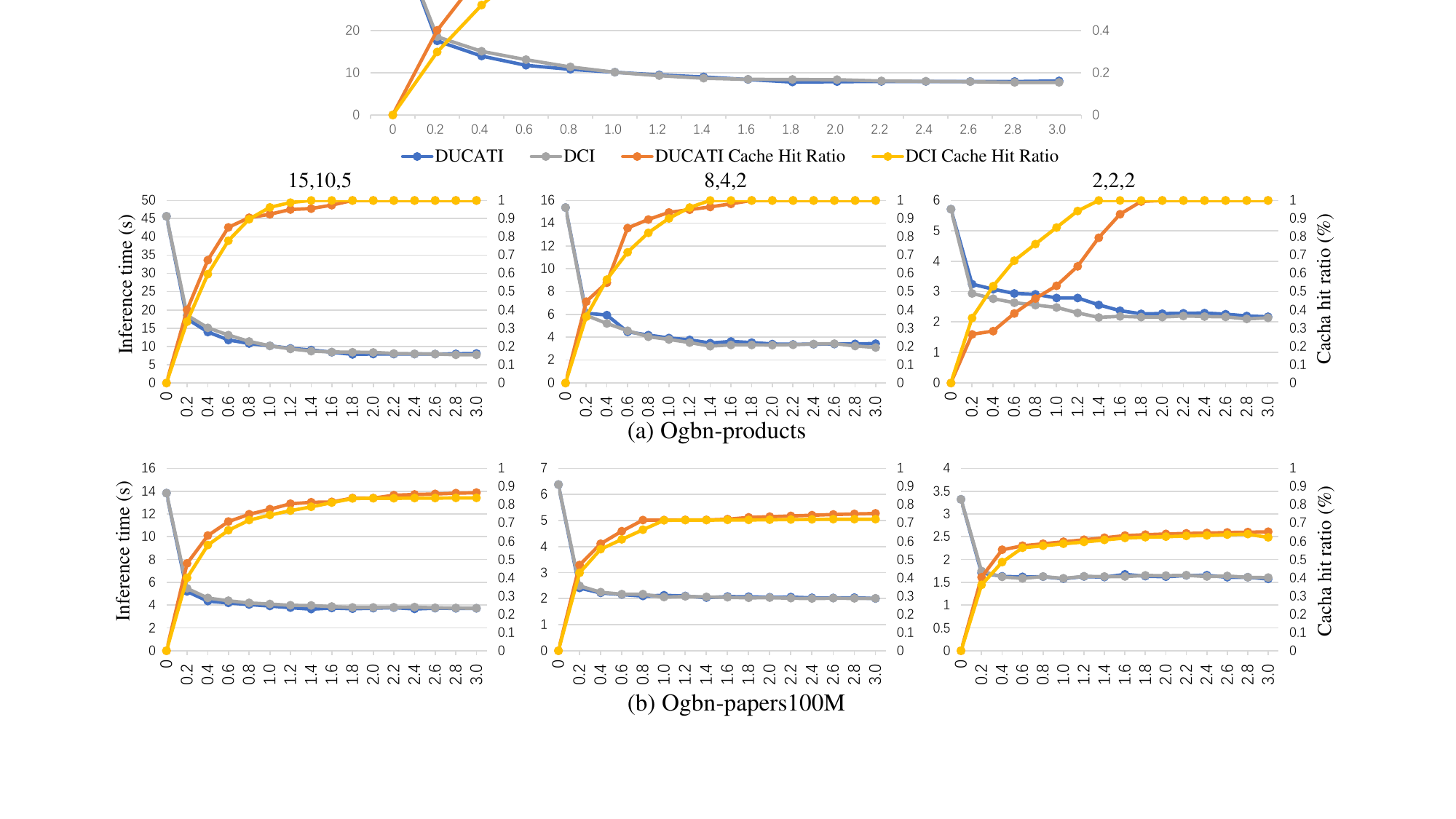}	
	\caption{Comparative analysis of inference speed and cache hit ratios for DCI and DUCATI algorithms across different fan-out (X-axis: cache capacity in GB).} 	
	\label{fig10}
\end{figure*}

Although preprocessing time was excluded when comparing DCI and DGL, inference on industrial-scale graphs can often surpass training in terms of time consumption. This is because the training set typically constitutes only a small portion of the overall dataset~\cite{zhu2024glisp}. Additionally, in real-world applications such as recommendation systems and fraud detection, graph structures and features are continuously updated. The trained model frequently performs inference on these updated graphs, leading to inference workloads that far exceed those of training. Given that resource-intensive preprocessing tasks consume significant computational resources, the preprocessing time of DCI, RAIN, and DUCATI will be compared.

\textbf{DCI vs RAIN.} The preprocessing time of DCI and RAIN was first compared, using the same experimental parameters as in the previous section. The comparison results are shown in Table~\ref{tab4}. In the majority of cases, DCI’s preprocessing time is less than 10\% of RAIN’s, and it never exceeds 47\%, even in the most demanding scenarios. On average, this time is merely 13.01\% of what is observed for RAIN. In summary, DCI significantly reduces the time required for preprocessing, demonstrating the efficiency of the algorithm.

\begin{figure}[t]	
	\centering	
	\includegraphics[width=0.5\textwidth, trim=11.2cm 4cm 8.6cm 6.5cm, clip]{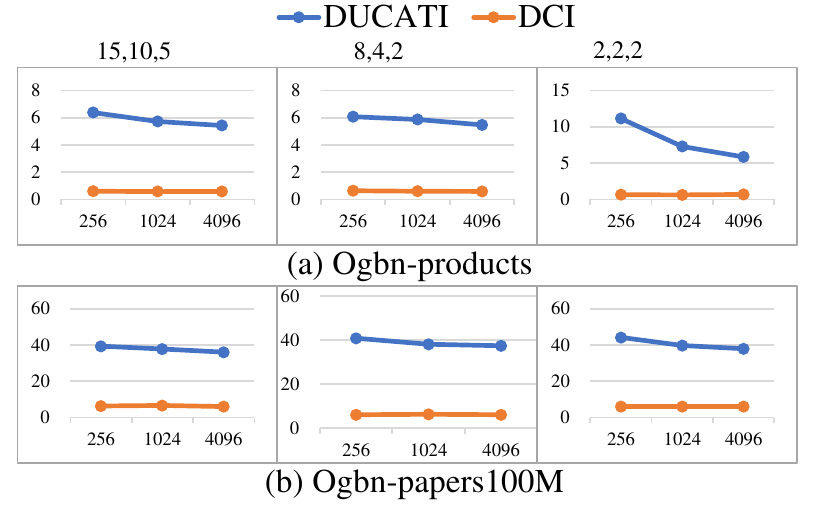}	
	\caption{Preprocessing time for DCI and DUCATI under different parameters(Y-axis unit: seconds, X-axis: batch size)} 	
	\label{figa5}
\end{figure}

\textbf{DCI vs DUCATI.} DUCATI, a dual-cache system primarily designed for training, was adapted for integration into DCI by isolating and incorporating its cache allocation and filling algorithms, replacing DCI's algorithms. The ogbn-products and ogbn-papers100M datasets were utilized, chosen for their real-world analogous size and structure, with results displayed in Fig.~\ref{figa5}. Comparative analyses revealed significant reductions in DCI’s preprocessing times—88.91\% to 94.37\% (average 90.49\%) on ogbn-products and 81.38\% to 84.95\% (average 82.81\%) on ogbn-papers100M. This improvement is attributed to DUCATI’s robust, training-focused cache allocation method, which includes analyzing value curves of 'nfeat' and 'adj' entries, determining slopes through curve fitting, and employing a knapsack-like strategy for cache allocation. Although feasible in training through amortization across epochs, this approach proves impractical during inference. In contrast, DCI optimizes computational and cache efficiencies by leveraging hot nodes and workload during pre-sampling, markedly reducing preprocessing times. The somewhat diminished performance on the ogbn-papers100M dataset is due to the substantial overhead from Unified Virtual Addressing (UVA) generation, exacerbated by the dataset's extensive size.

Additionally, by analyzing Fig.~\ref{figa5} and Table~\ref{tab4}, it is found that the preprocessing overhead of DCI is minimal, dependent solely on the number of preprocessing batches and the fan-out strategy used. In contrast, the RAIN algorithm employs Locality-Sensitive Hashing (LSH) to cluster similar batches, which results in a linear time complexity of \(O(n)\). Meanwhile, DUCATI adopts a knapsack-like problem-solving approach, featuring a time complexity of \(O(n \log n)\). In the subsequent sections, data on cache hit rates under various parameters will be presented to corroborate the reliability and efficiency of the algorithm.

\subsection{Cache Strategy Analysis: DCI and DUCATI}

To thoroughly evaluate the dual-cache systems DCI and DUCATI, and validate the effectiveness of the cache allocation and dual-cache filling algorithms, additional comparative experiments were conducted. The total cache budget was determined based on the method recommended by DUCATI. Specifically, the DGL inference system was run without caching to observe memory usage across different configurations, thereby determining the total cache capacity. A notable observation is that when the cache capacity is large enough to accommodate the entire dataset on the GPU, the performance of both strategies is identical, as all adjacency matrices and node features are cached, eliminating any performance differences due to different allocation strategies. Therefore, scenarios were set up to simulate the impact of both strategies on total runtime under GPU memory constraints, assuming available GPU memory ranging from 0GB to 3GB. The results are presented in Fig.~\ref{fig10}.

Overall, while there are some differences in the allocation of cache capacity between DCI and DUCATI, the average runtime difference between the two is less than 4\%. In some cases, DCI's strategy even outperforms DUCATI's strategy. This is because, under smaller fan-out, DUCATI may not fit the optimal slope in preprocessing. For the ogbn-products dataset, both DCI and DUCATI strategies achieve a 100\% cache hit rate once the total cache budget exceeds 2GB, as this is sufficient to cache the entire dataset on the GPU, leading both caching strategies to achieve the same inference speed ultimately. In contrast, as shown in Fig.~\ref{figa1}, the single-cache system stabilizes the feature loading time once the node feature budget surpasses 1GB, highlighting a key limitation of single-cache systems—allocating all available GPU memory to node features does not fully utilize the GPU memory. DCI allocates part of the memory to the adjacency matrix, thereby accelerating the sampling process and achieving better GPU memory utilization. For the ogbn-papers100M dataset, both strategies tend to allocate more cache to node features, and since this dataset follows a power-law distribution—where a few high-frequency samples dominate while numerous low-frequency samples contribute minimally—high cache hit rates are achieved after caching only a small portion of high-frequency samples. A common phenomenon observed across both datasets is that larger fan-out result in higher cache hit rates. This is because larger fan-out are more likely to capture high-frequency samples.

\begin{figure}[htbp]	
	\centering	
	\includegraphics[width=0.5\textwidth, trim=11.2cm 7.2cm 9.5cm 4cm, clip]{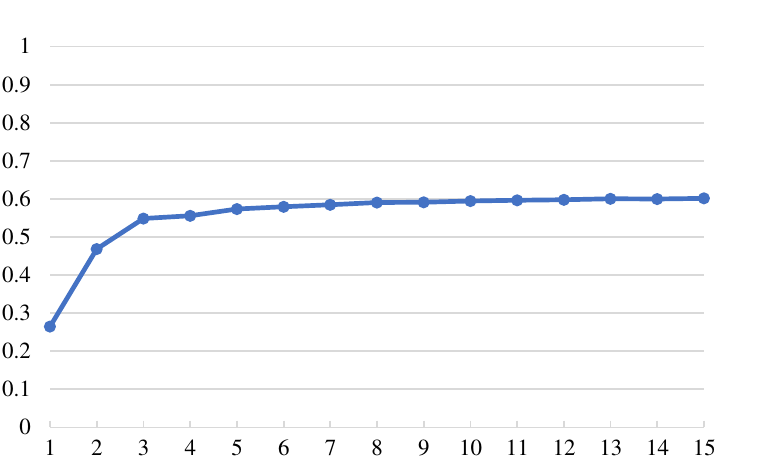}	
	\caption{Impact of Different Numbers of Preprocessing Mini-Batches on Cache Hit Rates (Y-axis unit: Cache hit rates, X-axis: Numbers of preprocessing mini-batches)} 	
	\label{fig11}
\end{figure}

Finally, the impact of the number of preprocessing batches on cache hit rates under conditions of limited cache capacity was also examined. The ogbn-products dataset was chosen for this test, with the total cache capacity set to 0.4GB. As can be seen from Fig.~\ref{fig10}., the cache hit rate does not reach 100\% when the total cache capacity is set to 0.4GB, allowing us to clearly observe the impact of different numbers of preprocessing rounds on cache hit rates. The experimental results, as shown in Fig.~\ref{fig11}, indicate that cache hit rates tend to stabilize when the number of preprocessing batches exceeds eight. Previously, systems targeting training typically chose epochs as the unit for preprocessing. The experiments demonstrate that using mini-batches as the unit can still achieve desirable hit rates, thus proving that DCI can achieve good results through rapid preprocessing.

\section{Conclusion}
In this paper, DCI is proposed, an efficient dual-cache system specifically designed to accelerate GNN inference, featuring a lightweight cache capacity allocation and filling strategy tailored for inference applications. Workloads under various parameter settings were analyzed, revealing that the load of sampling and node feature loading during GNN inference is variable, and traditional single-feature cache systems fail to fully utilize hardware resources. Therefore, an adjacency matrix cache is introduced alongside the node feature cache, forming a dual-cache system. DCI dynamically allocates cache capacity based on workload characteristics and employs a lightweight cache-filling algorithm to minimize redundant data loading, thereby enhancing hardware resource utilization. Experimental results show that DCI accelerates sampling and node feature loading across various scenarios, achieving end-to-end inference speedups of \(1.18\times\) to \(11.26\times\) over DGL, \(1.14\times\) to \(13.68\times\) over RAIN, and an average speedup of \(1.14\times\) over the most advanced single-cache systems for GCN, and \(1.2\times\) for GraphSAGE. In terms of preprocessing time, DCI achieves a reduction of 52.8\% to 98.7\% compared to RAIN. Additionally, compared to DUCATI's dual-cache population algorithm, which also employs a dual-cache strategy, DCI's population algorithm achieved an average reduction of \(90.49\%\) in preprocessing time on the ogbn-products dataset and \(82.81\%\) on the ogbn-papers100M dataset, while maintaining nearly the same inference performance.

\section*{Acknowledgment}
This work is supported financially by * Natural Science Foundation for Distinguished Young Scholar (ID: *), National Natural Science Foundation (ID: *).

\ifCLASSOPTIONcaptionsoff
  \newpage
\fi



%\bibliographystyle{IEEEtran}
%\begin{thebibliography}{1}
%\end{thebibliography}

\bibliographystyle{IEEEtran}  % 使用 IEEE 样式
\bibliography{aa.bib}  % 替换为你的 .bib 文件名（不带扩展名）

%\begin{biography}[{\includegraphics[width=1in,height=1.25in,clip,keepaspectratio]{mshell}}]{Michael Shell}

%\newpage


%\vfill


\end{document}